\documentclass[prd,aps,reprint,superscriptaddress,nofootinbib]{revtex4-2}

\pdfoutput=1

\usepackage[colorlinks=true
,urlcolor=blue
,anchorcolor=blue
,citecolor=blue
,filecolor=blue
,linkcolor=blue
,menucolor=blue
,linktocpage=true
,pdfproducer=medialab
,pdfa=true
]{hyperref}
\usepackage[T1]{fontenc}
\usepackage{fontawesome}
\usepackage{feynmf}
\usepackage{graphicx}
\usepackage{enumitem}
\usepackage{latexsym}
\usepackage{amsfonts}
\usepackage{amssymb}
\usepackage{mathrsfs}
\usepackage{color}
\usepackage{amsmath}
\usepackage[capitalise]{cleveref}
\usepackage{slashed}
\usepackage{dcolumn}
\usepackage{verbatim}
\usepackage{comment}
\usepackage{float}
\usepackage{multirow}
\usepackage{xspace}
\usepackage[normalem]{ulem}
\usepackage[dvipsnames]{xcolor}

\def\triumf{TRIUMF, 4004 Wesbrook Mall, Vancouver, BC V6T 2A3, Canada}
\def\sfu{Department of Physics, Simon Fraser University, Burnaby, BC V5A 1S6, Canada}
\def\uvic{Department of Physics and Astronomy, University of Victoria, Victoria, BC V8P 5C2, Canada}

\begin{document}

\title{Characterizing Dark Bosons at Chiral Belle}
\affiliation{\triumf}
\affiliation{\uvic}
\affiliation{\sfu}

\author{Carlos Henrique de Lima}
\email{cdelima@triumf.ca}
\affiliation{\triumf}

\author{David McKeen}
\email{mckeen@triumf.ca}
\affiliation{\triumf}

\author{Afif Omar}
\email{aomar@triumf.ca}
\affiliation{\triumf}
\affiliation{\uvic}

\author{Douglas Tuckler}
\email{dtuckler@triumf.ca}
\affiliation{\triumf}
\affiliation{\sfu}

\begin{abstract}
We explore the advantages of a polarized electron beam at Belle II, as proposed for ``Chiral Belle'', in the search for invisibly decaying (dark) bosons that weakly couple to the Standard Model. By measuring the polarization dependence of the production cross section of dark bosons in association with a photon, the dark boson's spin and Lorentz structure of its couplings can potentially be determined. We analyze the mono-photon channel, $e^+ e^- \rightarrow \gamma + \text{invisible}$, in detail, focusing on the production of an on-shell spin-1 boson. We explore this in the context of three separate scenarios for a new dark vector: a dark photon, a mass-mixed ``dark $Z$'', and a vector that couples to right-handed electrons, and estimate how well the couplings of such bosons to electrons can be constrained in the event of a positive signal.
\end{abstract}

\maketitle

%%%%%%%%%%%%%%%%%%%%%%
\section{Introduction}
%%%%%%%%%%%%%%%%%%%%%%

In defiance of its many successes, there is strong reason to believe that the Standard Model (SM) of particle physics is not the last word when it comes to a fundamental understanding of nature. One of the most pressing deficiencies of the SM is its lack of a candidate for the dark matter observed in the Universe. There is overwhelming evidence for dark matter astrophysically and cosmologically~\cite{Salucci:2018hqu,*Hamilton:2000du,*Bahcall_2003,*Allen_2011,*Simon_2019}, while thus far there has been no conclusive sign of dark matter in Earth-based searches.

Dark matter (DM) so far has only been probed gravitationally, and little is known about its fundamental nature, in particular, its mass and non-gravitational interactions. Dark matter candidates have been proposed over a wide range of masses, from $\sim 10^{-20}~\rm eV$ to the Planck scale or even larger if the dark matter is a macroscopic object~\cite{Bozorgnia:2024pwk,*Lin:2019uvt,*Lisanti:2016jxe,*Bertone:2004pz,*Jungman:1995df}. Given this ignorance, it is important that we cast a wide net in our search for DM and thoroughly explore the mass scales that we have access to experimentally.

In recent years, DM with a mass between about $10~\rm MeV$ and a few $\rm GeV$ has received increased attention~\cite{Essig:2013lka,*Alexander:2016aln,*Battaglieri:2017aum,*Agrawal:2021dbo,*Gori:2022vri,*deLima:2020lzl}. In this mass range, DM can be difficult to detect in direct detection experiments even if it is not very feebly coupled to the SM~\cite{Schumann:2019eaa}. A simple setup to realize such a scenario involves DM coupled to the SM through the vector or scalar portals -- that is, DM is coupled to a vector or scalar boson that weakly interacts with SM fields, which we generically term a dark boson. The existence of such a mediator is also motivated to evade the Lee-Weinberg lower bound on sub-GeV DM~\cite{Lee:1977ua}. If the dark boson is produced at a terrestrial experiment through its small coupling to the SM and its mass is more than twice the DM mass, it is natural that the mediator decays to DM and appears experimentally as missing momentum. 

In this work, we focus on the production of such invisibly decaying bosons in $e^+ e^-$ collisions at the Belle II experiment~\cite{Belle-II:2010dht}. Belle II has an edge in searches for this type of DM mediator since the center-of-mass energy of $10.58~\rm GeV$ offers a relatively large mass reach, while the fact that electrons and positrons are fundamental particles means that the scattering kinematics are well constrained and missing momentum signatures can be fully reconstructed.

Recently, the possibility of polarizing the $e^-$ beam at Belle II has begun to be explored~\cite{Liptak:2021opc,Roney:2022jdx,USBelleIIGroup:2022qro}, a project referred to as Chiral Belle. This would allow SM parity violation to be probed at a mass scale that would shed light on results from KTeV~\cite{NuTeV:2001whx}, to determine the vector couplings to the $Z$ of all charged leptons, $c$, and $b$ quarks with precision comparable to $Z$ pole measurements~\cite{Roney:2025lwo}, to measure the anomalous magnetic moment of the $\tau$ lepton with a precision of $\sim10^{-6}$ to $10^{-5}$~\cite{Bernabeu:2007rr,Crivellin:2021spu,Gogniat:2025eom}, improve the prevision on the Michel parameters in $\tau$ decay~\cite{RoneyEEFACT}, and access to the dynamical component of
the jet mass~\cite{Accardi:2022oog}, further elucidating the hadronization mechanism. Chiral Belle has a strong motivation, namely, extending our understanding of the Standard Model. At the same time, it can be used to explore the fundamental nature of dark sectors in some well-motivated scenarios.

Dark bosons at Belle II have been studied extensively for unpolarized beams~\cite{Essig:2013vha, Belle-II:2018jsg,Wakai_2021,Graham:2021ggy,Hearty:2022wij,Batell:2022dpx,Acanfora:2023gzr,Bauer:2023loq, Corona:2024xsg} . In these studies, it is seen that the expected reach for dark bosons in particular is world-leading over a large mass range, largely because of the incredible integrated luminosity of $50\ \text{ab}^{-1}$ that Belle II is expected to reach by the end of its running. Recently, it was pointed out that it is possible to distinguish the spin of a dark boson with Chiral Belle~\cite{Bauer:2023loq}. 

We extend the study of dark bosons further, where we highlight the potential to determine the Lorentz structure of the boson's coupling to electrons at Chiral Belle. Because we assume the boson's decay products go undetected, we cannot make use of their angular distributions to determine the Lorentz structure of the boson's couplings as in Ref.~\cite{Lee:2020tpn}. We therefore rely on the polarization dependence of the production cross section to distinguish the different Lorentz structures.

In this work, we focus on light dark vectors ($<10$ GeV) which decay on-shell to invisible states. We work out the polarization dependence of the production of an arbitrary dark boson, $X$, in the mono-photon channel, $e^+ e^- \rightarrow \gamma X$. While a dark (pseudo-)scalar has no polarization dependence, as shown in Appendix~\ref{sec:ap}, the vector has a non-trivial dependence on beam polarization in the presence of both vector and axial-vector couplings. This analysis is further extended for off-shell decays and heavier masses in~\cite{ZmonoBelle}. 

We highlight that the choice of beam polarization dictates which type of new physics the experiment is most sensitive to. Our study considers electron beam polarizations of $\pm70\%$, in line with current Chiral Belle proposals~\cite{USBelleIIGroup:2022qro}. We further show how the ability to separate the data according to $e^-$ polarization opens up the possibility of determining the Lorentz structure of such a dark vector in the event of a positive signal. Our study helps identify the major detector-related limitations to searches of this type, notably gaps in the detector coverage and photon reconstruction inefficiencies.

This paper is organized as follows. In Sec.~\ref{sec:1}, we introduce the generic dark vector mediator and discuss scenarios that motivate different vector and axial vector couplings. In Sec.~\ref{sec:2}, we discuss both the $e^+ e^- \rightarrow \gamma X$ process and the indirect probe from $e^+ e^- \rightarrow X^* \rightarrow f \bar{f} $, where $f$ represents a SM fermion, to measure the Lorentz structure with a polarized beam. Section~\ref{sec:3}  introduces the proposed search of the mono-photon channel at Chiral Belle. We highlight that most features of the background are polarization independent and thus most of the machinery of the unpolarized search can be reused. We conclude in Section~\ref{sec:conc}. In Appendix~\ref{sec:ap}, we discuss the scalar mediator, which is insensitive to beam polarization.

%%%%%%%%%%%%%%%%%%%%%%%%%%%%%%
\section{A Dark Vector}
%%%%%%%%%%%%%%%%%%%%%%%%%%%%%%
\label{sec:1}
We begin our analysis by considering a spin-1 boson, $V$, with mass $m_V$ and generic vector and axial-vector couplings to electrons,
\begin{equation}
\mathcal{L}
\supset
-V_\mu\bar e\gamma^\mu\bigl(g_V + g_A\gamma^5\bigr)e.
\label{eq:VLagrangian}
\end{equation}
We are agnostic about the generation of the boson's mass, which could be done with either the Higgs or Stueckelberg mechanisms. We further assume that $V$ couples to a neutral, collider-stable state $\chi$ (that could be dark matter) with mass less than half $m_V$, and that $V$ decays dominantly into $\chi$ pairs. The parameter region where $V$ is off-shell is similar to the standard model $Z$ process to neutrinos, which is explored further in~\cite{ZmonoBelle}.

Below, we discuss different realizations of specific models that can be described by the generic interactions in Eq.~(\ref{eq:VLagrangian}).

%%%%%%%%%%%%%%%%%%%%%%%%%%%
\subsection{Kinetically mixed dark photon}
\label{sec:dp}
%%%%%%%%%%%%%%%%%%%%%%%%%%%

A particularly simple way to couple a vector boson to the SM is to kinetically mix it with the $U(1)$ hypercharge gauge boson~\cite{Okun:1982xi,*Galison:1983pa,*Holdom:1985ag}
\begin{equation}
    \mathcal{L}\supset \frac{\epsilon}{2\cos\theta_W}F^\prime_{\mu\nu}B^{\mu\nu},
\end{equation}
where $B^{\mu\nu}$ and $F^\prime_{\mu\nu}$ are the field strengths of hypercharge and the new vector, respectively, $\epsilon$ is the kinetic mixing parameter, and $\theta_W$ is the weak mixing angle. When $m_V\ll m_Z$, as is the case when it can be produced at Belle II, it inherits couplings through the kinetic mixing that are primarily photon-like (i.e. vectorial) and is then often dubbed the ``dark photon''. However, it does mix slightly with the $Z$, which has chiral couplings to the SM fermions. Specifically, its couplings to the electron in this mass regime are~\cite{Curtin:2014cca}
\begin{equation}\label{eq:DPcoups}
\begin{aligned}
g_V\simeq-\epsilon e,~g_A&\simeq -\frac14\frac{m_{V}^2}{m_Z^2\cos^2\theta_W}\epsilon e
\\
&\simeq10^{-3}\left(\frac{m_V}{5~\rm GeV}\right)^2g_V.
\end{aligned}
\end{equation}
Thus, we see that in the dark photon case, if its mass is such that it can be produced on-shell at a $B$-factory, the spin-1 boson's couplings are almost vectorial with the fractional chirality at the $10^{-3}$ level or below.

%%%%%%%%%%%%%%%%%%%%%%%%%%%
\subsection{Mass mixed dark $Z$}
\label{sec:zd}
%%%%%%%%%%%%%%%%%%%%%%%%%%%

Another possibility to weakly couple a new massive spin-1 boson to the SM is to allow for mass mixing with the SM $Z$ boson, a scenario that leads to a ``dark $Z$''~\cite{Davoudiasl:2012ag}. The relevant term in the Lagrangian is
\begin{equation}
\begin{aligned}
{\cal L}\supset-2\epsilon_Zm_Z^2Z_d^\mu Z_\mu.
\end{aligned}
\end{equation}
where $\epsilon_Z$ is a small mixing parameter that can be generated in, e.g., an extended Higgs sector. In the two-Higgs doublet-model of~\cite{Davoudiasl:2012ag}, $\epsilon_Z\simeq \delta\times m_V/m_Z$ with $\delta$ a parameter of the theory. Atomic parity violation (APV) measurements limit $\delta\lesssim10^{-2}$~\cite{Marciano:1990dp,*Porsev:2010de}. In this scenario, the dark $Z$  boson couplings to SM fermions are $Z$-like, and, in particular for the electron, are given by
\begin{equation}\label{eq:DZcoups}
\begin{aligned}
g_V&=\frac{e}{\sin2\theta_W}\left(-\frac12+2\sin^2\theta_W\right)\epsilon_Z\simeq -0.05\, \epsilon_Z e,\\
g_A&=-\frac{e}{\sin2\theta_W}\left(-\frac12\right)\epsilon_Z\simeq 0.6\, \epsilon_Z e\simeq -13g_V,
\end{aligned}
\end{equation}
using $\sin^2\theta_W\simeq0.23$~\cite{ParticleDataGroup:2024cfk}. In this case, the fractional chirality of the couplings to the electron is at the $10^{-1}$ level, a much larger value than in the dark photon case.

%%%%%%%%%%%%%%%%%%%%%%%%%%%
\subsection{Right-handed vector}
\label{sec:vr}
%%%%%%%%%%%%%%%%%%%%%%%%%%%

Vectors that couple to right-handed SM fermions have been considered to explain a number of experimental anomalies~\cite{Batell:2011qq,Karshenboim:2014tka}. Generally, such vectors require additional fermions to render the model theoretically self-consistent, potentially leading to other signals~\cite{Preskill:1990fr,*Batra:2005rh,*Dobrescu:2014fca}. Moreover, since these vectors couple to non-conserved currents at low scales, the emission of longitudinally polarized vectors is enhanced and can lead to very stringent constraints~\cite{Dror:2017ehi,Dror:2017nsg} although there is strong model dependence on the anomaly-canceling UV content of the theory~\cite{Gan:2020aco}. In Ref.~\cite{Dror:2017nsg}, it was shown that a limit on the branching ratio for $B^+\to K^++{\rm inv.}$ of $2.5\times10^{-5}$ restricts
\begin{equation}
\begin{aligned}
\left|g_V+g_A\right|&\lesssim 6\times10^{-6}\left(\frac{m_V}{\rm GeV}\right),
\end{aligned}
\end{equation}
for $m_V<m_B-m_K$ in the case of an invisibly decaying vector coupled universally to right-handed SM fermions~\cite{Kahn:2016vjr}. There is now a measurement from Belle II of the branching fraction for this mode of $(2.3\pm0.7)\times10^{-5}$~\cite{Belle-II:2023esi} which differs at a more than $2\sigma$ from the SM prediction; for a discussion of potential explanations of this see, e.g.,~\cite{Altmannshofer:2023hkn,*McKeen:2023uzo,*Fridell:2023ssf,*He:2024iju,*Buras:2024ewl,*Altmannshofer:2024kxb,*Bhattacharya:2024clv,*Calibbi:2025rpx}. For $m_V<m_K-m_\pi\simeq 350~{\rm MeV}$, the limit from kaon decays on the size of the coupling is stronger by about an order of magnitude.

Similar to the dark $Z$ scenario, APV in such a model limits
\begin{equation}
\begin{aligned}
\left|g_V+g_A\right|&\lesssim 1\times10^{-4}\left(\frac{m_V}{\rm GeV}\right),
\end{aligned}
\end{equation}
roughly at the same level as in the dark $Z$ scenario above.

It is worth noting that gauging left-handed charged leptons is significantly more challenging. It generally leads to enhanced interactions of neutrinos with matter, which are heavily constrained experimentally~\cite{Batell:2011qq}. Additionally, our study below will focus on the coupling of the new boson to electrons and not make use of the model-dependent limits above that rely on additional couplings to quarks.

%%%%%%%%%%%%%%%%%%%%%%%%%%%
\subsection{Generic dark vector}
\label{sec:gen}
%%%%%%%%%%%%%%%%%%%%%%%%%%%

It is also natural to consider new vectors with couplings that are a linear combination of the benchmark values discussed above. In particular, it is difficult to forbid kinetic mixing for symmetry reasons when introducing a new vector associated with a $U(1)$ gauge theory. Depending on the particulars of the UV completion, the resulting photon-like couplings can be more or less relevant than the original couplings that were induced.

For example, in the $Z_d$ scenario, a loop-induced kinetic mixing is generated at low scales, which is of order
\begin{equation}
\begin{aligned}
\epsilon\sim \frac{e}{16\pi^2}\sum_f g_V^f Q_f\sim 10^{-3}\epsilon_Z,
\end{aligned}
\end{equation}
where we have assumed that $f$ runs over all SM fermions. While this is generally small in comparison to $\epsilon_Z$, it is possible that in the UV, interactions that generate $\epsilon_Z$ are accompanied by interactions that generate a comparable $\epsilon$. In this case, the low energy coupling to electrons of $V$ would be roughly
\begin{equation}
\begin{aligned}
g_V&\simeq- \epsilon e,\\
g_A&\simeq 0.6\, \epsilon_Z e\simeq -0.6\times\frac{\epsilon_Z}{\epsilon}\times g_V,
\end{aligned}
\end{equation}
with $\epsilon_Z/\epsilon$ determined by the UV theory. 

With these models in mind, we now focus on the search for dark vectors at Belle with arbitrary $g_V$ and $g_A$ couplings to electrons.

%%%%%%%%%%%%%%%%%%%%%%%%%%%%%%%%%%
\section{Dark Vectors at (Chiral) Belle II}
%%%%%%%%%%%%%%%%%%%%%%%%%%%%%%%%%%
\label{sec:2}

At an $e^+e^-$ collider, if $m_V$ is less than the center-of-mass energy $\sqrt s$, the leading signature is a mono-photon signal, $e^+e^-\to\gamma V$, with a photon recoiling against an invisibly decaying $V$. This signal has been considered extensively in the literature in the case of unpolarized $e^\pm$ beams~\cite{Borodatchenkova:2005ct,Batell:2009yf,Essig:2009nc,BaBar:2017tiz}. In this work, we are primarily interested in what we can learn if measurements are made with a nonzero polarization of the electron beam as envisioned in the Chiral Belle proposal.

For a right- or left-polarized electron beam (and the positron beam unpolarized), the $e^+e^-\to\gamma V$ cross section as a function of the center-of-mass angle between the photon and electron, $\theta_\gamma^\ast$, is
\begin{equation}
\begin{aligned}
\frac{d\sigma_{R,L}}{d\cos\theta_\gamma^\ast}&=\left(\frac{g_V\pm g_A}{e}\right)^2\bar\sigma_{\gamma V},
\end{aligned}
\label{eq:sigVLR}
\end{equation}
where $e=\sqrt{4\pi\alpha}$ is the photon's coupling strength and
\begin{equation}
\begin{aligned}
\bar\sigma_{\gamma V}&=\frac{2\pi\alpha^2}{s}\times\frac{2}{\sqrt s}\frac{{E_V^\ast}^2+{E_\gamma^\ast}^2\cos^2\theta_\gamma^\ast}{{E_\gamma^\ast}\sin^2\theta_\gamma^\ast}
\\
&= 1.25~{\rm nb}\times\frac{2}{\sqrt s}\frac{{E_V^\ast}^2+{E_\gamma^\ast}^2\cos^2\theta_\gamma^\ast}{{E_\gamma^\ast}\sin^2\theta_\gamma^\ast},
\end{aligned}
\label{eq:sigbar}
\end{equation}
where in the final equality we have normalized on the Belle II center-of-mass energy $\sqrt s=10.58~\rm GeV$ with $\alpha(\sqrt s)\simeq 1/132$. The center-of-mass energies of the photon and new boson, $E_\gamma^\ast$ and $E_V^\ast$ respectively, are given by
\begin{equation}
\begin{aligned}
E_\gamma^\ast=\frac{\sqrt s}{2}\left(1-\frac{m_V^2}{s}\right),~E_V^\ast=\frac{\sqrt s}{2}\left(1+\frac{m_V^2}{s}\right).
\end{aligned}
\label{eq:EgammaEV}
\end{equation}

It is clear from the expression in Eq.~(\ref{eq:sigVLR}) that the cross section depends on the electron polarization when both $g_V$ and $g_A$ are nonzero -- in other words, when the left- and right-handed electrons have different couplings to the new gauge boson. To describe an electron beam with generic polarization, we define
\begin{equation}
\begin{aligned}
P=\frac{L^--R^-}{L^-+R^-}=2L^--1=1-2R^-,
\end{aligned}
\label{eq:P}
\end{equation}
where $R^-$ ($L^-$) is the fraction of electrons that are right-(left-)polarized and $R^-+L^-=1$. When scattering on a beam of unpolarized positrons, the cross section to produce $\gamma V$ is
\begin{equation}
\begin{aligned}
    \sigma (P) &= \left(\frac{g_V^2}{e^2} + \frac{g_A^2}{e^2} - 2 P \frac{g_V g_A}{e^2} \right)\bar\sigma_{\gamma V}
    \\
    &\equiv\epsilon_{\rm eff}^2\times\bar\sigma_{\gamma V}\, ,
    \label{eq:sigmaP}
\end{aligned}
\end{equation}
where we have defined an effective coupling parameter $\epsilon_{\rm eff}$.

In addition to the direct production signal, $V$ can also mediate $s$-channel SM fermion pair production in a manner similar to the SM $Z$ boson. If $V$ is chirally coupled, then this can lead to a difference between the production cross sections for each polarization. Considering the process $e^+e^-\to f\bar f$, where $f$ is a SM fermion other than the electron, and supposing further that $V$ couples to $f$ analogously, the fractional difference between the $e^-$-polarized $f\bar f$ production cross sections is
\begin{equation}
\begin{aligned}
A_{LR}^{f\bar f}\equiv\frac{\sigma_L^{f\bar f}-\sigma_R^{f\bar f}}{\sigma_L^{f\bar f}+\sigma_R^{f\bar f}}\simeq\frac{s}{m_V^2-s}\frac{2}{Q_f}\frac{g_A^e g_V^f}{e^2}\times\frac{P_L-P_R}{2},
\end{aligned}
\end{equation}
where $P_{L,R}$ is the polarization defined in Eq.~(\ref{eq:P}) for the nominally left- or right-polarized $e^-$ beam and we have used superscripts to specify the couplings of $V$ to the electron and $f$, writing the electric charge of $f$ as $Q_f$ where $Q_e=-1$. (When the couplings and mass are rescaled appropriately, this matches the well-known SM value involving $Z$ exchange in, e.g., Ref.~\cite{Roney:2022jdx}.)

We can estimate the reach on $g_V\times g_A$ from measurements of $A_{LR}^{f\bar f}$ that will be made at Chiral Belle. The precision on $A_{LR}^{\mu^+\mu^-}$ from $e^+e^-\to \mu^+\mu^-$ is estimated to be at the level of $10^{-5}$~\cite{Aleksejevs:2018bzh}. Therefore, if the dark vector couples universally to charged leptons ($g_{V,A}^e=g_{V,A}^\mu\equiv g_{V,A}$), then this measurement could constrain
\begin{equation}
\begin{aligned}
\frac{\sqrt{g_A g_V}}{e}\lesssim 2\times10^{-3},
\end{aligned}
\end{equation}
for $m_V=5~\rm GeV$. Interestingly, the sign of the asymmetry changes as $m_V$ crosses the center-of-mass energy. Similar considerations apply to other SM final states if the dark vector couples to them.

More model-independently, we can constrain just the dark vector coupling to electrons  through the asymmetry in Bhabha scattering, $e^+e^-\to e^+e^-$. In this case, the presence of $t$-channel exchange in addition to $s$-channel complicates the expression for $A_{LR}^{e^+e^-}$ and introduces a dependence on the scattering angle. Making use of estimates of the sensitivity to $A_{LR}^{e^+e^-}$ in this channel in Ref.~\cite{Miller:2024ivn} leads to a future constraint of roughly
\begin{align}
    \frac{\sqrt{g_A g_V}}{e}\lesssim 3\times10^{-3},
\end{align}
comparable to that from muons (if the muon couples in the same way as the electron). 

In the next section, we will see that these bounds are not competitive with direct searches for $V$ in the mono-photon channel, when $V$ is light enough to be directly produced. However, when $m_V$ is such that it is only produced off-shell, these constraints can be important.

%----------------------------------------------------------------------------------------
\begin{figure*}[t!]
\includegraphics[width=0.43\linewidth]{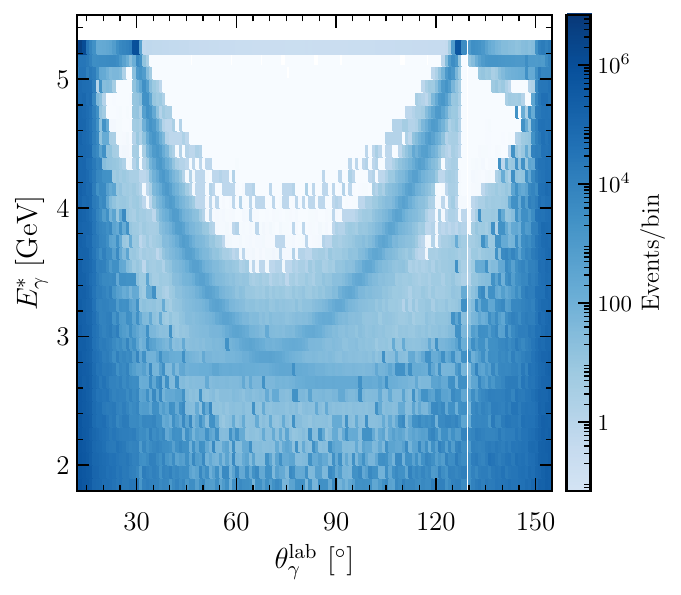}
\includegraphics[width=0.43\linewidth]{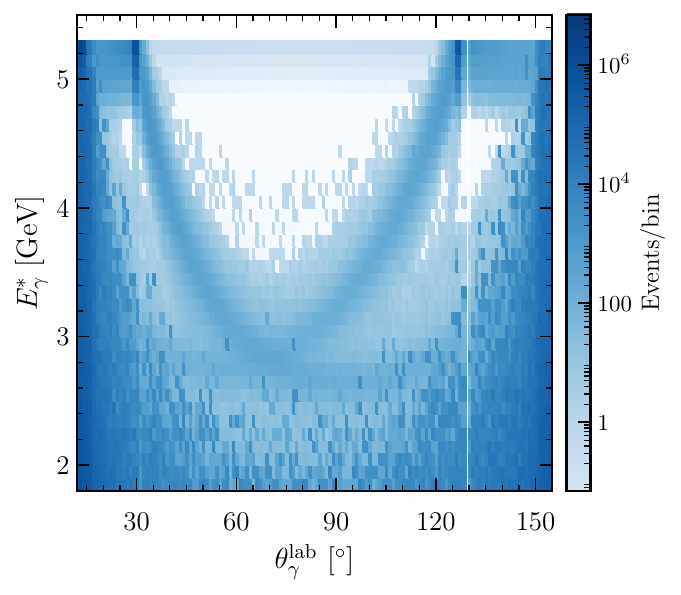}
\caption{Truth-level (\textbf{left}) and smeared (\textbf{right}) distribution of the mono-photon SM background for 20~fb$^{-1}$ of data as a function of the center-of-mass energy of the measured single photon $E_\gamma^\ast$, and the photon lab-frame angle $\theta_\gamma^\text{lab}$ relative to the electron beam, with the ECL geometry information included. The arch shapes are a consequence of the endcap gaps~(which are visible in the histogram) that give a significantly higher probability of missing a particle there and still measuring a photon in the ECL. The concentration of events at high energies comes from $\gamma \gamma$ events where one photon is missed inside the main barrel of the ECL.}
\label{fig:hist}
\end{figure*}
%----------------------------------------------------------------------------------------
%%%%%%%%%%%%%%%%%%%%%%%%%%%%%%%%%%%%%%%%%%%%%%%%%%%%%%%%%%%%%%%%%%%%%
\section{Mono-photon search at (Chiral) Belle II}
\label{sec:3}
%%%%%%%%%%%%%%%%%%%%%%%%%%%%%%%%%%%%%%%%%%%%%%%%%%%%%%%%%%%%%%%%%%%%%
In this section, we study the sensitivity of (Chiral) Belle II to an invisibly decaying vector in the mono-photon channel. We consider several realistic values of the electron polarization and integrated luminosity at Chiral Belle to determine the distinguishing power that a polarized beam brings to this search.

%%%%%%%%%%%%%%%%%%
\subsection{Analysis}
%%%%%%%%%%%%%%%%%%
\label{sec:4}

We perform our analysis for both an unpolarized and a $\pm70\%$ polarized electron beam. In both the unpolarized and polarized cases, the signal for our analysis is a single hard photon, as we assume that the dark vector decays invisibly. To avoid inefficiencies at lower photon energies, we consider only photons with center of mass energy $E_{\gamma}^{*}>1.8$ GeV. The dominant backgrounds are electromagnetic and independent of the electron beam polarization at leading order. They are
\begin{itemize}
    \item $e^+e^- \rightarrow e^+e^- \gamma(\gamma)$ where the final state $e^\pm$ are undetected,
    \item $e^+e^- \rightarrow \gamma\gamma(\gamma)$ where only one photon is observed.
\end{itemize}
There is also a smaller, irreducible electroweak background which does depend on the electron polarization, namely,
\begin{itemize}
    \item $e^+e^- \rightarrow \nu_{l}\bar{\nu}_{l}\gamma(\gamma)$.
\end{itemize}
This consists of the same final state as the signal: a single photon recoiling against missing momentum. The only difference between this and the signal process is the virtuality carried by the $\nu_{l}\bar{\nu}_{l}$ pair. This background thus features a range of center-of-mass energies carried by the photon, in contrast to a monochromatic photon as in the signal case (in the center-of-mass frame). This irreducible background is an interesting electroweak physics target in its own right, and is explored further in~\cite{ZmonoBelle}.

There is also a smaller background, that predominantly affects the low-energy photon spectrum, from $e^+e^-\rightarrow q\bar{q}(\gamma)$ or, more generically, $e^+e^-\rightarrow\text{Hadrons}\, +\gamma$ and also $e^+e^- \rightarrow \tau^+ \tau^- \gamma$. The final-state photon can originate from different sources: either from the decay of final-state hadrons, from initial-state emission, or from emission from final-state charged particles. Most of these events contain additional tracks and can be vetoed in mono-photon searches. A small contribution arises from charged particles outside the tracking region, which then become irreducible backgrounds.

Finally, there is an irreducible background originating from photon detection inefficiencies when one of the photons goes  missing inside the detector. For this analysis, the dominant contribution comes from $e^+e^- \rightarrow \gamma\gamma$,\footnote{This inefficiency also significantly affects other channels used to search for invisibly decaying bosons, e.g. $e^+ e^- + \text{invisible}$~\cite{Acanfora:2023gzr}, where $e^+ e^- \gamma$ becomes an irreducible background.} where one photon is undetected. We assume a photon reconstruction inefficiency of $10^{-6}$, which closely resembles current Belle II capabilities~\cite{Belle-II:2018jsg,HeartyGod}. The challenging part of this background is that it predominantly affects the high-energy photon region, which no other QED process populates. Reducing this inefficiency would significantly enhance Belle II's reach in invisible searches.

%----------------------------------------------------------------------------------------
\begin{figure*}[t!]
\includegraphics[width=\linewidth]{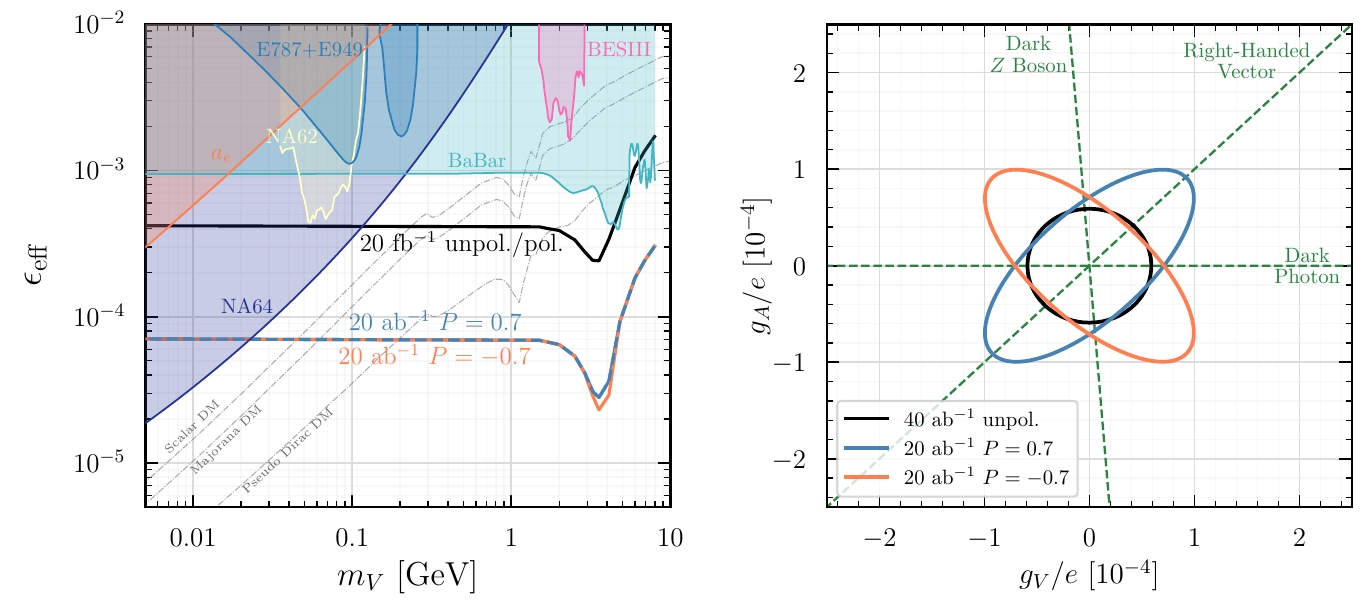}
\caption{\textbf{Left:} Estimated Belle II 95\% CL upper limits on $\epsilon_{\rm eff}$ as a function of $m_V$ given the analysis described in the text, for both unpolarized and polarized beams with a luminosity 20~fb$^{-1}$ (black curve), a polarized electron beam with $P = 0.7$ and a luminosity of 20~ab$^{-1}$ (blue curve), and a polarized electron beam with $P=-0.7$ and a luminosity of 20~ab$^{-1}$ (orange curve). The bound from polarized beams should be interpreted as being on $\epsilon_{\rm eff}$ as defined in Eq.~(\ref{eq:sigmaP}). Existing limits for invisible dark photon searches from  E787/E949~\cite{Essig:2013vha,E787:2001urh,*BNL-E949:2009dza,*Davoudiasl:2014kua}, BaBar~\cite{BaBar:2017tiz}, NA62~\cite{NA62:2019meo}, BESIII~\cite{BESIII:2022oww}, NA64~\cite{NA64:2023wbi}, and the electron anomalous magnetic moment $a_e$~\cite{Fan:2022eto} (assuming a vector coupling and the measurement of $\alpha$ in Ref.~\cite{Morel:2020dww}---taking an axial-vector coupling and the larger $\alpha$ in Ref.~\cite{Parker:2018vye} does not change the bound substantially).  The gray-dashed curves are thermal DM targets with $m_V = 3 m_{\rm DM}$ and $\alpha_D = 0.5$ for scalar, Majorana, and pseudo-Dirac DM candidates~\cite{Berlin:2018bsc}. 
%--------------------------------------------------------------------------------
\textbf{Right:} The 95\% CL limits in the $g_V$ vs $g_A$ plane at Chiral Belle for $m_V = 100$ MeV, assuming 20~ab$^{-1}$ of $70\%$ left-handed (orange) or  $70\%$ right-handed (blue) polarized electron beams (regions outside the ellipses are excluded). We also show the Belle II reach for unpolarized beams with 40~ab$^{-1}$ of data (black). The dashed green lines indicate the relationship between $g_A$ and $g_V$ for a dark photon, a dark $Z$ boson, and a right-handed vector as labeled.}
\label{fig:bounds}
\end{figure*}
%----------------------------------------------------------------------------------------

For a photon to be reconstructed, we require that it deposit energy in the electromagnetic calorimeter (ECL). The ECL angular acceptance, including both endcaps and main barrel, is defined by $12.4^\circ < \theta < 31.4^\circ $, $32.2^\circ < \theta < 128.7^\circ$, and $130.7^\circ < \theta < 155.1^\circ$, where $\theta$ is the lab-frame angle with respect to the initial $e^-$ beam~\cite{Adachi:2018qme}. At Belle II, efficient photon reconstruction is only available in the region $18.5^\circ < \theta < 139.2^\circ$, and we focus our analysis on signal photons well inside this region~\cite{Belle-II:2018jsg}.

We consider a charged particle to be lost if it carries transverse momentum less than $p_T^{\rm min}=0.2~\rm GeV$, or does not leave a track (which we take to mean that its lab-frame scattering angle is not in the range $17^\circ < \theta_{e^\pm} ^{\text{lab}}<150^\circ$). An effect that we do not consider in our analysis is the contamination of the signal region from cosmic rays. This can potentially weaken the expected discovery reach as discussed in~\cite{Wakai_2021}.

We generate the background events for both unpolarized and polarized $e^-$ beam at leading order in QED using \texttt{MadGraph5\_aMC@NLO}~\cite{Alwall:2014hca} interfaced with \texttt{Pythia8.2}~\cite{Sjostrand:2014zea} for showering and hadronization. Because QED processes have large cross sections at this energy, we generate 90 million Monte Carlo events to ensure that we do not miss any relevant kinematic features. In Fig.~\ref{fig:hist}, we show the distribution of the backgrounds with $20~{\rm fb}^{-1}$ of data in the $E_\gamma^\ast$ vs $\theta^\text{lab}_\gamma$ plane.

Noteworthy features of Fig.~\ref{fig:hist} are the arch-like shapes at intermediate values of $\theta_\gamma^{\rm lab}$ and $E_\gamma^\ast$. This shape is due to the interplay of the Belle II detector geometry and three-body final states. This was verified by calculating the three particle phase space distribution with a flat matrix element and including the gaps in the detector. Thus, this background can only be reduced if no gaps are present between the main barrel and the endcaps. The other important feature of the distribution is the concentration of events at half of the center-of-mass energy. These originate from the $\gamma \gamma$ backgrounds where one of the photons is lost in the main barrel because of inefficiencies. Reduction of this background is possible with further improvements in the detector and reconstruction. 

Signal events have a unique photon center-of-mass energy, given in Eq.~(\ref{eq:EgammaEV}), and the lab-frame angular distribution can be related at leading order to the center-of-mass distribution in Eq.~(\ref{eq:sigbar}). The angular distributions are insensitive to polarization and exhibit the same shape in $\theta^\text{lab}_\gamma$ as the $\nu_{l}\bar{\nu}_{l}$ background process. Polarization effects enter only through the overall normalization of the cross section in both processes, while the shape is determined by the vector nature of the interaction.

For each mass hypothesis, we construct a region in $E_{\gamma}^{*}$ that contains $95\%$ of the signal events $s$. Given $b$ background events, we then calculate the optimal cut in $\theta_\gamma^{\rm lab}$ which maximizes the exclusion significance $Z_{\rm excl} =\sqrt{2(s-b\log(1+s/b))}$~\cite{Cowan:2010js}. Conservatively, we consider one single cut for each mass hypothesis, effectively choosing a single bin in $E_{\gamma}^{*}$ for each statistical test. This, combined with the high statistics of the Monte Carlo samples, helps avoid overfitting around statistical fluctuations. This approach is similar to that used in~\cite{Belle-II:2018jsg,Corona:2024xsg,Bauer:2023loq,Wakai_2021}.

When polarization effects are included, the only affected background is the subdominant neutrino pair production process. Signal extraction is also modified, as the polarized signal cross section is given by Eq.~(\ref{eq:sigmaP}) and proportional to $\epsilon_{\rm eff}^2=(g_V^2+g_A^2-2 P g_V g_A)/e^2$, where $ P$ is defined in Eq.~(\ref{eq:P}).

%%%%%%%%%%%%%%%%
\begin{figure*}[t!]
\includegraphics[width=0.325\linewidth]{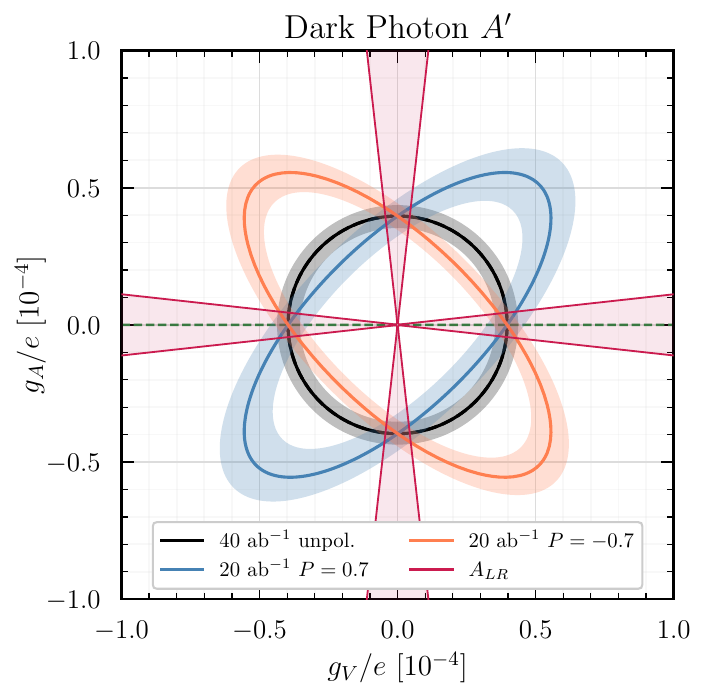}
\includegraphics[width=0.325\linewidth]{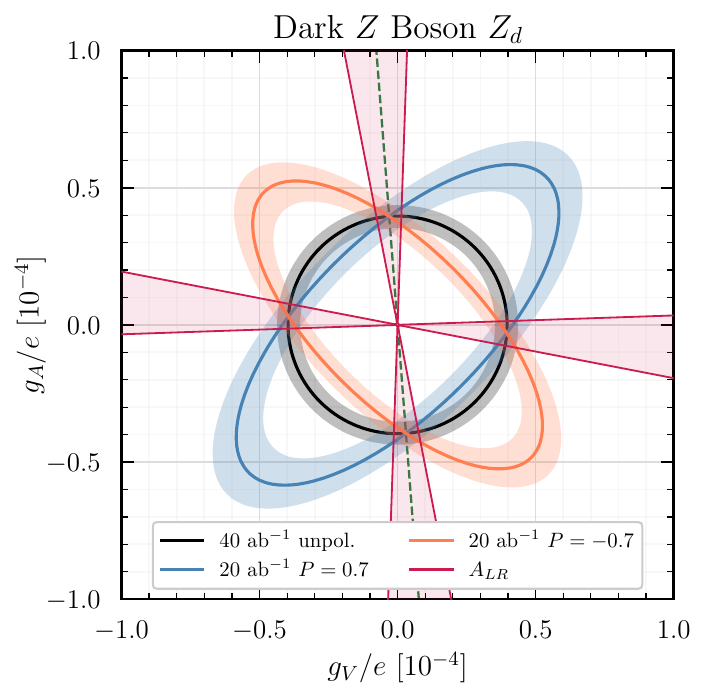}
\includegraphics[width=0.325\linewidth]{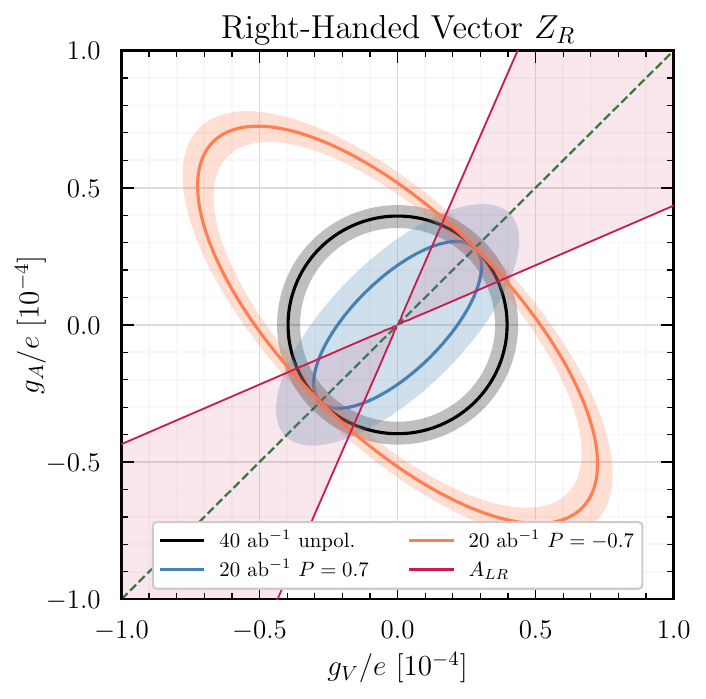}
\caption{Example of the discriminatory power from $e^-$ beam polarization for the three different models we consider. In these plots, we assume a future $5\sigma$ discovery with $40~{\rm ab}^{-1}$ of unpolarized data for $m_V=3.67~\rm GeV$, which carves out the gray circular $1\sigma$ preferred coupling regions. The $20~{\rm ab}^{-1}$ of $P=0.7$ and $P=-0.7$ (orange) data in this sample then require at $1\sigma$ that the couplings are in the blue and orange regions respectively, (partially) breaking the degeneracy in the $g_V$ and $g_A$ plane. The dashed green lines are the coupling relationships for each scenario. The red shaded regions show the couplings preferred at the $1\sigma$ level from measurements of $A_{LR}$ defined in Eq.~(\ref{eq:ALR}), which is less subject to systematic errors. Note that there is a degeneracy in the $A^\prime$ and $Z_d$ cases between (nearly) purely vector or axial-vector couplings so that the cross sections for left- or right-polarized $e^-$ beams are nearly the same. In the $Z_R$ case, the chiral nature of the couplings breaks this degeneracy.}
\label{fig:discovery}
\end{figure*}
%%%%%%%%%%%%%%%%%%%%%%%

We perform the statistical analysis under three scenarios. First, we consider only the geometric features of the detector. Second, we include energy smearing of the events, estimated with a Gaussian, to simulate the energy resolution of the Belle II ECL.\footnote{In the energy region of interest, the energy resolution is approximately constant, $\sigma_E/E = 0.0225$~\cite{Adachi:2018qme}, which we adopt as our smearing factor.} Third, we account for the inefficiency of photon detection. It is instructive to evaluate the reach of the mono-photon search at each of these stages of detector simulation, as described below.

We first consider an ideal detector with $20~\rm fb^{-1}$ of unpolarized data and a $100$ MeV dark vector. The signal region is localized at $E_{\gamma}^{*}=5.29~\rm GeV$, and we construct an energy bin from $4.89~\rm GeV$ to $5.29~\rm GeV$. The optimal cut that maximizes the $95\%$ CL exclusion sensitivity is $\theta_\gamma^{\rm lab}=[ 38^\circ, 115^\circ ]$ and inside this region we expect no background events. This translates into an estimated 95\% CL limit $\epsilon_{\rm eff} < 1.68 \times 10^{-4}$. Once we include smearing based on the detector energy resolution, events from lower energies bleed into the high energy region. We obtain a slightly weakened limit of $\epsilon_{\rm eff} < 1.74 \times 10^{-4}$. Finally, including a photon detection inefficiency of $10^{-6}$ in the main barrel of the ECL does not affect the optimal signal region but significantly weakens the limit to $\epsilon_{\rm eff} < 4.20 \times 10^{-4}$.

We show the estimated reach in coupling as a function of the dark vector's mass in the left plot of Fig.~\ref{fig:bounds}, folding in the effects of finite detector resolution and inefficiency as described above. We consider both unpolarized and polarized electron beams (with $P=\pm0.7$) and integrated luminosities of $20~\rm fb^{-1}$ and $20~\rm ab^{-1}$, and present our limits in terms of $\epsilon_{\rm eff}$ to account for measurements sensitive to different polarizations. Constraints from other measurements and regions of parameter space that can explain the DM abundance through freeze-out~\cite{Battaglieri:2017aum} are also shown.

The reach of Belle II with $20~\rm fb^{-1}$ of unpolarized or polarized data is depicted by the black curve. Note that the reach does not depend on the polarization since the neutrino background is not significant with this luminosity. We see that Belle II covers new parameter space compared to BaBar for $m_V \lesssim 3$ GeV. The reach for $P = 0.7$ with $20~\rm ab^{-1}$ of data is shown in the left plot of Fig.~\ref{fig:bounds} by the blue curve, while the reach for $P=-0.7$ is shown by the dashed orange curve. In both cases, dark vector couplings $\lesssim 10^{-4}$ will be probed over a broad range of dark vector masses. We see that, for most of the mass range, the reach for either polarization is essentially equal. This is because the dominant background at corresponding photon energies is due to QED and thus not sensitive to polarization. For $3~{\rm GeV}\lesssim m_V\lesssim 4~{\rm GeV}$, the subdominant neutrino background, which does depend on polarization, becomes important and leads to differences in the reach for each polarization. The usefulness of polarization information in actually measuring this SM process is further explored in Ref.~\cite{ZmonoBelle}.

\subsection{Model comparison} 
Using the analysis described above, the right panel of Fig.~\ref{fig:bounds} shows the upper bounds for a $100~\rm MeV$ boson in the $g_A$ vs $g_V$ plane assuming $20~{\rm ab}^{-1}$ of $P=0.7$ (blue) and $P=-0.7$ (orange) collisions which corresponds to $40~{\rm ab}^{-1}$ of unpolarized data (black). We also show the relationship between $g_V$ and $g_A$ in the three benchmark models discussed above by the dashed green lines. 

To show the distinguishing power of a polarized analysis, we consider the scenario of a future discovery with $m_V=3.67~\rm GeV$. We take the couplings such that $40~{\rm ab}^{-1}$ of unpolarized data show a $5\sigma$ excess compared to the background-only hypothesis. This corresponds to $\epsilon_{\rm eff}=(3.96\pm 0.4)\times10^{-4}$. We show the resulting $\pm1\sigma$ favored region in the $g_V$-$g_A$ plane in Fig.~\ref{fig:discovery} as the gray shaded circular regions for each of the three benchmark models we consider. Each of these models implies a different expected number of events in the two $20~{\rm ab}^{-1}$ of data for each $e^-$ polarization. We show the $\pm1\sigma$ expected couplings that would be extracted from the two polarization measurements as blue ($P=0.7$) and orange ($P=-0.7$) shaded regions. We see that the overlap of these analyses occurs in regions of the $g_A$ vs. $g_V$ plane that correspond to the assumed underlying models (cf. right panel of Fig.~\ref{fig:bounds}). 

When confronting real data, a convenient measure of parity violation that is less sensitive to systematic errors is the fractional asymmetry of polarized $\gamma V$ production,
\begin{equation}
A_{LR}\equiv\frac{\sigma_L-\sigma_R}{\sigma_L+\sigma_R}=-\frac{2 g_V g_A}{g_V^2+ g_A^2}P,
\label{eq:ALR}
\end{equation}
where we have assumed running with equal polarization magnitudes $\left|P\right|$ for left- and right-handed electron beams. In the $g_V$-$g_A$ plane, measuring this fractional asymmetry defines straight lines that constrain the model. In other words, the fractional asymmetry relates $g_A$ to $g_V$ but does not fix their overall scale. We show in red the regions that the measurement of $A_{LR}$ would carve out, assuming a positive signal as above and folding in the expected statistical errors, in each of the three models we consider in Fig.~\ref{fig:discovery}.

Because the dark photon and dark $Z$ cases both have approximately a single parity of coupling to the electron, $g_A\ll g_V$ and $g_V\ll g_A$ respectively, the regions of overlap are near the axes in the $g_V$-$g_A$ plane. In contrast, the $Z_R$ case leads to a large difference in the number of signal events between the two polarizations and can be distinguished from the previous two models. This can also hold for more generic dark vectors, possibly in cases where kinetic and mass mixing are comparable in importance.

%%%%%%%%%%%%%%%%%%%%%%
\section{Summary and Outlook}
\label{sec:conc}
%%%%%%%%%%%%%%%%%%%%%%

We have studied the prospects for detecting light, invisibly decaying dark vectors at Chiral Belle via the mono-photon process $e^+ e^- \rightarrow \gamma X$. Our work shows how polarized electron beams can play an important role in distinguishing the Lorentz structure of the dark sector. While unpolarized searches will offer world-leading sensitivity for this class of dark sector models, beam polarization opens a new window into the nature of the interactions.  

Our results build on and complement the already world-leading capabilities of Belle II for this class of dark matter models. The experiment’s high luminosity, clean environment, and ability to trigger on mono-photon events make it an ideal laboratory for precision measurements and rare searches. With the addition of polarization of the electron beam, Chiral Belle can go significantly further, not just in extending our understanding of the Standard Model, but opening the possibility to discern the fundamental structure of new dark states.

Our work highlights the main experimental limitations for invisible searches at Belle. The two main sources of irreducible backgrounds come from the detector geometry and detector/reconstruction inefficiencies, with the possibility of improving the latter. The fact that most backgrounds originate from QED and are polarization-independent may be leveraged in the event of a signal, which could have a generic chiral structure.

A straightforward extension of this work involves searches for heavier dark states that are produced off-shell at a $B$ factory. These scenarios are also well motivated, and there is overlap with  $e^+e^-\to \gamma Z^\ast\to\gamma\nu\bar\nu$, which itself could present interesting physics. We explore this scenario in detail in a follow-up work~\cite{ZmonoBelle}. Such searches can be complementary to direct~\cite{Curtin:2014cca,ATLAS:2024zxk,*CMS:2024jyb} or indirect~\cite{Hook:2010tw,*Kribs:2020vyk,*Thomas:2021lub,*Thomas:2022qhj,*Yan:2022npz} searches at higher energy colliders.

%\vspace{0.5cm}

\acknowledgments
We thank Michael Roney and Christopher Hearty for helpful discussions and feedback on the draft. This work is supported by Discovery Grants from the Natural Sciences and Engineering Research Council of Canada (NSERC). TRIUMF receives federal funding via a contribution agreement with the National Research Council (NRC) of Canada. This work was performed in part at the Aspen Center for Physics, which is supported by a grant from the Simons Foundation (1161654, Troyer) and National Science Foundation grant PHY-2210452. For facilitating portions of this research, DT wishes to acknowledge the Center for Theoretical Underground Physics and Related Areas (CETUP$^\ast$), the Institute for Underground Science at Sanford Underground Research Facility (SURF), and the South Dakota Science and Technology Authority for hospitality and financial support.
%\vspace{0.5cm}
\appendix

%%%%%%%%%%%%%%%%%%%%%%%%%%%%%%%%%%%%%%%%%%%
\section{Spinless boson is insensitive to electron polarization}
%%%%%%%%%%%%%%%%%%%%%%%%%%%%%%%%%%%%%%%%%%
\label{sec:ap}

Consider a generic spin-0 boson coupled to electrons at dimension-4 through
\begin{equation}
{\cal L}\supset-\phi\bar e\left(g_S+ig_P\gamma^5\right)e,
\end{equation}
and to photons at dimension-5,
\begin{equation}
\begin{aligned}
{\cal L}\supset \frac{\phi}{\Lambda} \left(g_{\gamma\gamma} F^{\mu\nu} F_{\mu\nu} + \tilde g_{\gamma\gamma}F^{\mu\nu}\tilde F_{\mu\nu} \right).
\end{aligned}
\end{equation}
with $g_{s,P}$, $g_{\gamma\gamma}$, and $\tilde g_{\gamma\gamma}$ real. Ignoring the electron mass, the cross sections for $e^+e^-\to\gamma\phi$ are the same for right- or left-polarized electrons,
\begin{align}
\frac{d\sigma_R}{d\cos\theta_\gamma^\ast} &=\frac{d\sigma_L}{d\cos\theta_\gamma^\ast}\nonumber
\\
\quad&=\frac{\alpha}{2 s}\Bigg[\frac{\left(g_S^2+g_P^2\right)\left(1+m_\phi^4/s^2\right)}{\left(1-m_\phi^2/s\right)\sin^2\theta_\gamma^\ast}
\\
\quad\quad&+\frac{\left(g_{\gamma\gamma}^2+\tilde g_{\gamma\gamma}^2\right)s}{2\Lambda^2}\left(1-\frac{m_\phi^2}{s}\right)^3\left(1+\cos^2\theta_\gamma^\ast\right)\bigg].\nonumber
\end{align}

Unlike the spin-one case discussed in the main text, polarized electron beams do not probe the Lorentz structure of scalars coupled to electrons or photons. The angular distributions are different, which can be used to distinguish the spin of the dark state in case of a positive measurement, provided enough data is accumulated. Furthermore, the equality of the extracted polarization cross sections (within errors) would be a test of the scalar nature of the dark boson produced.

\bibliography{ref_CBelle}

\end{document}